# Heuristic Algorithm for Interpretation of Non-Atomic Categorical Attributes in Similarity-based Fuzzy Databases – Scalability Evaluation

M. Shahriar Hossain
*Department of Computer Science*
*Montana State University*
*Bozeman, MT 59715, USA*
mshossain@cs.montana.edu

Rafal A. Angryk
*Department of Computer Science*
*Montana State University*
*Bozeman, MT 59715, USA*
angryk@cs.montana.edu

*Abstract* - **In this work we are analyzing scalability of the heuristic algorithm we used in the past [1-4] to discover knowledge from multi-valued symbolic attributes in fuzzy databases. The non-atomic descriptors, characterizing a single attribute of a database record, are commonly used in fuzzy databases to reflect uncertainty about the recorded observation. In this paper, we present implementation details and scalability tests of the algorithm, which we developed to precisely interpret such non-atomic values and to transfer (i.e. defuzzify) the fuzzy tuples to the forms acceptable for many regular (i.e. atomic values based) data mining algorithms. Important advantages of our approach are: (1) its linear scalability, and (2) its unique capability of incorporating background knowledge, implicitly stored in the fuzzy database models in the form of fuzzy similarity hierarchy, into the interpretation/defuzzification process.**

## I. Introduction

With recent, aggressive growth of data mining applications and their increasing influence on competitiveness of multiple businesses, the margin of tolerance, allowing previously to simply disregard imprecise data when performing data analysis, decreases dynamically. Similarity-based fuzzy relational databases let the user reflect uncertainty about inserted information via the insertion of multiple descriptors in every column of the data table. At the same time, majority of currently available data mining algorithms allow the user to deal only with atomic attribute values, forcing many data-miners to disregard fuzzy records. In our opinion, successful future of fuzzy database models becomes strongly correlated with the development of consistent and time-efficient mechanisms allowing to mine the imprecise (i.e. fuzzy) data. In this work, we evaluate a simple, heuristic technique allowing consistent interpretation of the non-atomic values stored in the fuzzy databases. This process performs mapping (i.e. defuzzifying) of imprecise, fuzzy records (containing non-atomic descriptive values) into sets of atomic records, which can be analyzed using regular data mining techniques. Our approach shows to have nice, scalable properties, but it has a strong, heuristic character and therefore the term "defuzzification" used in this paper should be interpreted in context slightly broader than typically used in the fuzzy sets' community (i.e. transformation of a fuzzy set to a single, crisp value). In this work we will focus on problem of interpretation of multiple, categorical values by utilization of expert knowledge stored in the form of fuzzy similarity relation.

Section II of this paper describes the background of the work. We present analysis of the scalability of our defuzzification algorithm in section III. The paper is concluded in section IV.

## II. Background

### A. Fuzzy Database Model

There are two fundamental properties of fuzzy relational databases, proposed originally by Buckles and Petry [5-6] and extended further by Shenoi and Melton [7-8]: (1) utilization of non-atomic attribute values to characterize features of recorded entities we are not sure of, and (2) ability of processing the data based on the domain-specific and expert-specified fuzzy relations applied in place of traditional equivalence relations.

In fuzzy database model it is assumed that each of the attributes has its own fuzzy similarity table, which contains degrees of similarity between all values occurring for the particular attribute. Such tables can be also represented in the form of the similarity hierarchies, named by Zadeh [9] partition trees that show how the particular values merge together as we decrease similarity level, denoted usually by $\alpha$.

Example of fuzzy similarity table defined for domain COUNTRY, which preserves max-min transitivity property defined by Zadeh [9], is presented in TABLE I. Fig. 1 represents a partition tree reflecting the fuzzy similarity class defined in the TABLE 1.

TABLE I
FUZZY SIMILARITY TABLE FOR DOMAIN COUNTRY [15]

|         | Canada | USA | Mexico | Colom. | Venez. | Austral. | N.Zlnd. |
|---------|--------|-----|--------|--------|--------|----------|---------|
| Canada  | *1.0*  | *0.8* | *0.8* | *0.4*  | *0.4*  | *0.0*    | *0.0*   |
| USA     | *0.8*  | *1.0* | *0.8* | *0.4*  | *0.4*  | *0.0*    | *0.0*   |
| Mexico  | *0.8*  | *0.8* | *1.0* | *0.4*  | *0.4*  | *0.0*    | *0.0*   |
| Colomb. | *0.4*  | *0.4* | *0.4* | *1.0*  | *0.8*  | *0.0*    | *0.0*   |
| Venezu. | *0.4*  | *0.4* | *0.4* | *0.8*  | *1.0*  | *0.0*    | *0.0*   |
| Austral.| *0.0*  | *0.0* | *0.0* | *0.0*  | *0.0*  | *1.0*    | *0.8*   |
| N.Zlnd. | *0.0*  | *0.0* | *0.0* | *0.0*  | *0.0*  | *0.8*    | *1.0*   |

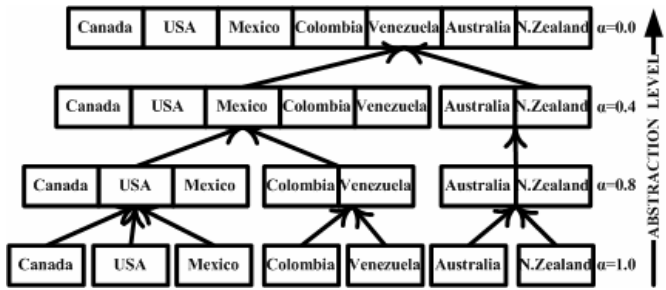

Fig. 1. Partition tree for domain COUNTRY derived from the TABLE I.

*B. Taxonomic Categorical Attributes and their Analysis*

In [10], chapter six has been devoted almost entirely to the problem of derivation of basic description statistic (in particular: medians, modes and histograms) from non-atomic values of categorical attributes. The authors propose transformation of non-atomic values to collection of pairs in the format of $(\xi,\varphi)$, where $\xi$ stands for a singleton (i.e. atomic) symbolic value, and $\varphi$ represents $\xi$'s observed frequency. The observed frequency distribution of a multi-valued variable $Z$, for a single data point, can be now defined as the list of all values in the finite domain of $Z$, together with the percentage of instances of each of the values in this data point (i.e. list of $\xi$ and $\varphi$ pairs). The authors [10] suggest that the observed frequency distribution may be related to the natural dependencies between the categorical values, as reflected by the provided attribute's taxonomy. This observation provided us with motivation for the work presented below.

*C. Similarity-driven Vote Distribution Method for Interpretation of Non-Atomic Categorical Values*

Based on the rationale presented above, we developed a simple, heuristic method to transfer/defuzzify non-atomic values in the fuzzy records to the collection of pairs including atomic values and their fractional observed frequencies. In our work, we focused on utilization of both: (1) the background knowledge about similarity of attributes' values, which is reflected in fuzzy databases by a predefined partition tree, and (2) the uncertain information, carried by collection of different values entered as data into a fuzzy record.

To explain our heuristic, we want a reader to assume for a moment that he/she needs to find a drugs' dealer who, as a not-confirmed report says (i.e. our fuzzy tuple), was recently seen in *{Canada, Colombia, Venezuela}*. The most trivial solution would be to split the count of observed frequency equally among all inserted descriptors, which is to interpret the entry as the following collection *{Canada|0.333, Colombia|0.333, Venezuela|0.333}*. This approach however does not take into consideration real life dependencies, which show that *Colombia* and *Venezuela* are closer to each other. Knowing that, we wanted these values to be emphasized a little more than *Canada*.

In our work [1-4], we replaced the even distribution of an individual fuzzy record's vote with a nonlinear spread, dependent both on the quantity of inserted values and on their similarity. Using the partition tree (Fig. 1), we can extract from the set of the originally inserted values those concepts which are more similar to each other than to the remaining values. We call them subsets of resemblances (e.g. *{Colombia, Venezuela}*). Then we use them as a basis for calculating a more fair (i.e. nonlinear) distribution of a fuzzy record's fractions. An important aspect of this approach is extraction of the subsets of resemblances at the lowest possible level of their common occurrence, since the nested character of fuzzy similarity relation guarantees that above this $\alpha$-level they are going to co-occur regularly.

TABLE II
FUZZY SIMILARITY TABLE FOR DOMAIN COUNTRY [11]

| OUTPUT | COMMENTS |
|---|---|
| {Canada, Colombia, Venezuela} | 0.0 | STORED |
| {Canada, Colombia, Venezuela} | ~~0.0~~ 0.4 | STORED, UPDATED |
| {Canada} | 0.8 | STORED |
| {Canada} | ~~0.8~~ 1.0 | STORED, UPDATED |
| {Colombia, Venezuela} | 0.8 | STORED |
| {Colombia} | 1.0 | STORED |
| {Venezuela} | 1.0 | STORED |

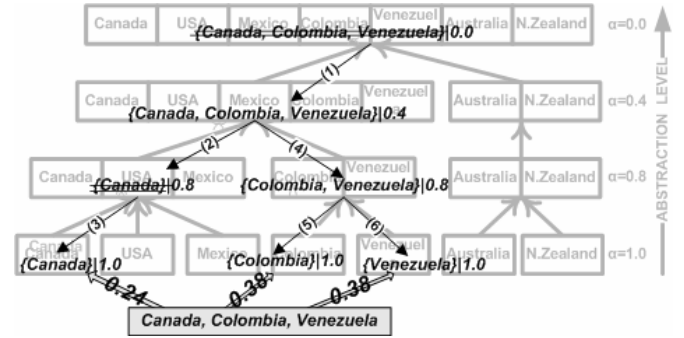

Fig. 2. Partition tree for domain COUNTRY derived from the TABLE I.

Given (1) a set of values inserted as a description of particular entity's attribute, and (2) a hierarchical data structure reflecting Zadeh's partition tree [9] for the attribute; we can extract a table, which includes (a) the list of all subsets of resemblances from the given set of descriptors, and (b) the highest level of $\alpha$-proximity of their common occurrence. We then can use the list (first column in TABLE II) to fairly distribute fuzzy record's fractions to the collection of fractional, precise records with atomic attribute values.

Our algorithm uses preorder recursive traversal for searching the partition tree. If any subset of the given set of descriptors occurs at the particular node of the concept hierarchy we store the values that were recognized as similar, and the corresponding value of $\alpha$. An example of such a search for subsets of resemblances in a tuple with the values *{Canada, Colombia, Venezuela}* is depicted in Fig. 2. Numbers on the links in the tree represent the order in which the particular subsets of similarities were extracted.

After extracting the *subsets of resemblances*, we apply a summarization of $\alpha$ values as a measure reflecting both the frequency of occurrence of the particular attribute values in the *subsets of resemblances*, as well as the abstraction level of these occurrences. Since during the search the country

*Canada* was reported only twice, we assigned it a grade *1.4* (i.e. *1.0+0.4*). For *Colombia* we get: *Colombia|(1.0 + 0.8 + 0.4) = Colombia|2.2*, and for the last value: *Venezuela|(1.0 + 0.8 + 0.4) = Venezuela|2.2*.

At the very end we normalize grades assigned to each of the entered values: *Canada |(1.4/5.8) = Canada |0.24, Colombia |(2.2/5.8) = Colombia |0.38, Venezuela |(2.2/5.8) = Venezuela |0.38*. This leads to the new distribution of the record's fractions, which, in our opinion, more accurately reflects real life dependencies than a linear-split approach.

## III. IMPLEMENTATION OF DEFUZZIFICATION ALGORITHM AND SCALABILITY TESTS

In this section we discuss implementation details of our algorithm, followed by experimental tests we conducted when analyzing its scalability. Our experiments have been conducted using some artificial datasets and different types of similarity trees.

### A. Defuzzification Algorithm

Our algorithm utilizes preorder recursive tree traversal (depth-first search, DFS) for finding matching subsets between a given fuzzy tuple and similarity classes in the partition tree. Our goal is to find the largest matching subset at each node of the partition tree. The high level outline of the algorithm is portrayed in TABLE III. The procedure *partitionTreeTraversal* is written in such a manner that it

TABLE III
DEFUZZIFICATION ALGORITHM

**Procedure**: *partitionTreeTraversal*
**Input Papameters:** *List sList, TreeNode masterNode, Vector searchVector, int level*
**Returen Type:** *Vector*
**Method:**
(1)   *List subsetList = sList*.clone();
(2)   *int totalChildren = masterNode*.getTotalChildren();
(3)   *Vector masterVector = masterNode*.getNodeVector();
(4)   *Vector subsetVector* = get the longest subset from *subsetList* that is found in the root of *masterNode*;
(5)   if (*subsetVector*.size()==0)
(6)      return *{Φ}*;
(7)   else if (*subsetVector*.size() = = *searchVector*.size())
(8)      update the corresponding entry for each element of *searchVector* or *subsetVector* with corresponding *α*–value of *level*–th level ;
(9)   else
(10)  add corresponding *α*-value of *level*–th level to the corresponding entry for each entity of *subsetVector*;
(11)  *Vector resultVector = searchVector*.clone();
(12)  for (*int i=0; i<totalChildren; i++*){
(13)     *Vector temp* = call *partitionTreeTraversal* with input parameters *subsetList, masterNode.getChild(i), subsetVector*, and *(level+1)*;
(12)     *subsetList = subsetList –* (all subsets that have at least one entity of *temp*);
(13)     *resultVector = resultVector – temp*;
(14)     if (*resultVector = = {Φ}* )   // NOTE 1
(15)        break ; }
(16)  if (*totalChildren= =0*)   // NOTE 2
(17)     return *subsetVector*;
NOTE 1 Do not traverse other branches because all entities are already found in the previously traversed branches.
NOTE 2 If *masterNode* is a leaf.

TABLE IV
DESCRIPTION OF *TreeNode* CLASS

**Class**: *TreeNode*
**Local Variables:**
 *int childCount;*
 *Vector nodeData, childrenNodes;*
**Initializatins:**
 *childCount = 0; nodeData = {Φ}; childrenNodes = {Φ};*
**Constructor:**
(1)   Public TreeNode (*Vector root*) {
(2)      *nodeData* = (*Vector*) *root*.clone( );
(3)      *childrenNodes* = new *Vector()*;
(4)   }
**Public Methods:**
(5)   public *void* addChild(*TreeNode childNode*){
(6)      *childrenNodes*.add(*childNode*);
(7)      *childCount*++; }
(8)   public *int* getTotalChildren( ){ // NOTE 1
(9)      return *childCount*; }
(10)  public *Vector* getNodeVector ( ){ // NOTE 2
(11)     return *nodeData*; }
(12)  public *void* setNodeAsLeaf ( ){
(13)     *childrenNodes*.clear( ); }
(14)  public *Vector getChildrenVector* ( ){// NOTE 3
(15)     return *childrenNodes*; }
(16)  public *TreeNode* getChild (*int i*){ // NOTE 4
(17)     return ( (*TreeNode*) *childrenNodes*.get(*i*) ); }
NOTE 1 e.g., in the case of a node with three descendents, this function would return 3.
NOTE 2 returns the parent node.
NOTE 3 returns the children nodes as a *Vector*.
NOTE 4 returns *i*-th child where each child itself is a *TreeNode*

avoids unnecessary branches while traversing the tree. There is no need to search a subtree, if there is no overlap between the values registered in the fuzzy tuple and the parent of this subtree. *List* and *TreeNode* parameters of the *partitionTreeTraversal* procedure are passed by reference where other parameters are passed by value. The method *clone()* of an object returns an entirely new instance of the cloned object to avoid changes to the referenced parameter, while the referenced parameter is necessary to be kept intact for the upper levels of recursion. A *Vector* can store a collection of entered attribute values, e.g. *{French, Greek, Italian}* and it can be used as an instance of *searchVector* of the algorithm. Besides, the *Vector* can also contain a collection of other *Vectors*. On the other hand, a *List* is a collection of *Vectors* e.g., during the first call of *partitionTreeTraversal* the parameter *sList* should contain a sorted (descending order, based on the size of subsets) list of all the possible subsets of *searchVector* except the empty subset (*{Φ}*). The operation denoted by "–" in the defuzzification algorithm is considered a regular *SetDifference* operation.

The implementation of our defuzzification system is based on a class named *TreeNode* (the class description is depicted in TABLE IV), which contains only a single node of the fuzzy similarity hierarchy, pointers to its immediate descendants, and some public methods to access them. Each descendant is an instance of the same class, called *TreeNode*. Therefore, the *TreeNode* contains pointers to instances of descendent *TreeNode*s. Descendents that do not point to any other child node point to null *TreeNode* (*{Φ}*).

As unnecessary branches are always omitted by the defuzzification algorithm, the performance of the algorithm depends on the distribution of the *searchVector* in the partition tree. Average computational cost is dominated by the probability of the existence of a *searchVector* in a certain node at certain level of the partition tree.

*B. Scalability Tests*

The experiments in this section are conducted using some artificial datasets and four different types of partition trees. The domain size of the experimental attribute is 32. Fig. 3 contains the partition trees we have used for our experiments. Fig. 3(a, b, c and d) have 2, 3, 4 and 5 levels of abstraction respectively. Fig. 3(d) also shows how the values are merged at higher levels. For simplicity, this representation has been omitted in Fig. 3(a, b and c).

Data points used in our analysis were generated by picking values randomly from the attribute domain, where the domain is defined as the set of all 32 values specified at the lowest (i.e. $\alpha=1.0$) level of the partition tree. We carried out the tests for two different degrees of imprecision reflected in our test data. Number of unique values for each fuzzy record (stored in *searchVector* during defuzzification) was picked in such way, that they would reflect 75% and 12.5% of imprecision respectively (i.e., for 75% imprecise data set, we randomly picked 24 different values out of 32-values long domain for each of the test records; 12.5%-imprecise data have been generated by randomly choosing only 4 descriptors). The number of fuzzy records in each imprecise data file was 30,000, so that the random characteristic could be evenly distributed among the partition tree as the defuzzification algorithm omits the unnecessary branches. Such large number of fuzzy records, containing an attribute with fixed level of imprecision, has been chosen to ensure a random (i.e., average)-case traversal, which we found the most informative when comparing our experiments.

We perform the scalability check with these two different percentages of imprecision: 75% and 12.5%. Fig. 4 and Fig. 5 depict the corresponding graphs of our scalability tests.

Fig. 4(a) shows that the defuzzification algorithm has linear performance with respect to number of fuzzy records with all the four types of partition trees. As *H4* has the highest number of levels in its corresponding partition tree, it takes the longest time when compared with other trees (it generates the topmost line of the graph of Fig. 4(a) ). In contrary, *H1* has only two levels and therefore takes the shortest amount of time at every corresponding number of records, if compared with other three hierarchies (*H2*, *H3* and *H4*). Fig. 4(b) shows average time to traverse partition tree for a single record with the same data as used to generate results presented in Fig. 4 (a). In Fig. 4(b), timing behaviors with all the partition trees used are expected to be linear and parallel to the horizontal axis. Beyond our expectations, it is evident from Fig. 4(b) that the average time of defuzzification of a record is slightly higher for low number of fuzzy records with a particular tree. It may lead to misleading conclusion that the algorithm performs better with large number of records. We would rather explain this odd behavior by usually constant data structure initialization costs, which typically have more significant influence on algorithm's average running time when considered in the context of smaller datasets.

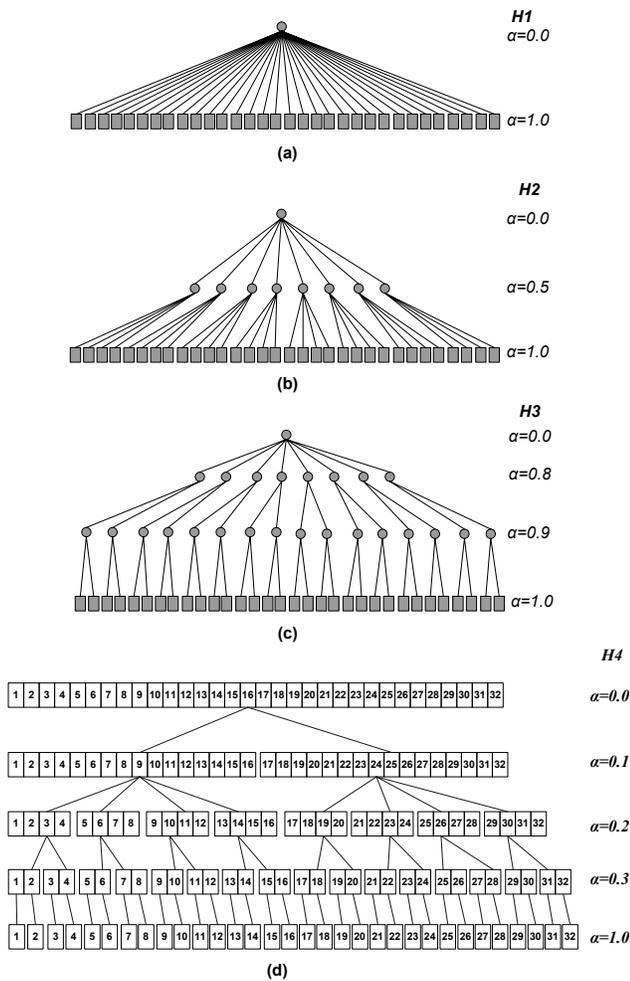

Fig. 3. Partition trees used for the scalability test. Corresponding $\alpha$–values are given at right side of the partition trees and labelled as *H1*, *H2*, *H3* and *H4*, correspondingly for the four partition trees.

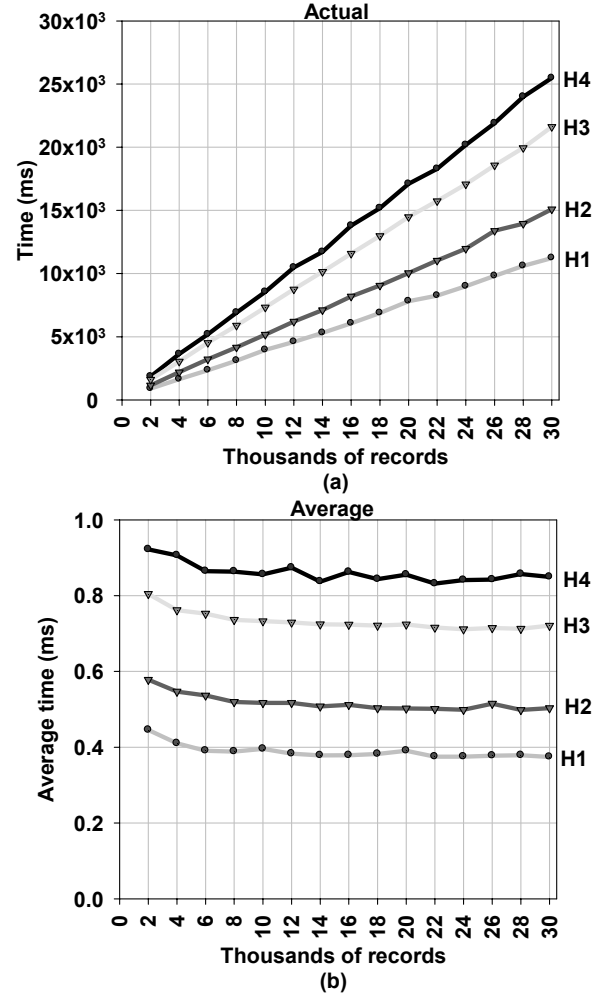

Fig. 4. Algorithm's scalability for 75% of data imprecision. **(a)** Actual time taken by our algorithm with partition trees *H1*, *H2*, *H3* and *H4*. **(b)** Average time to traverse a single record by the algorithm for the same trees.

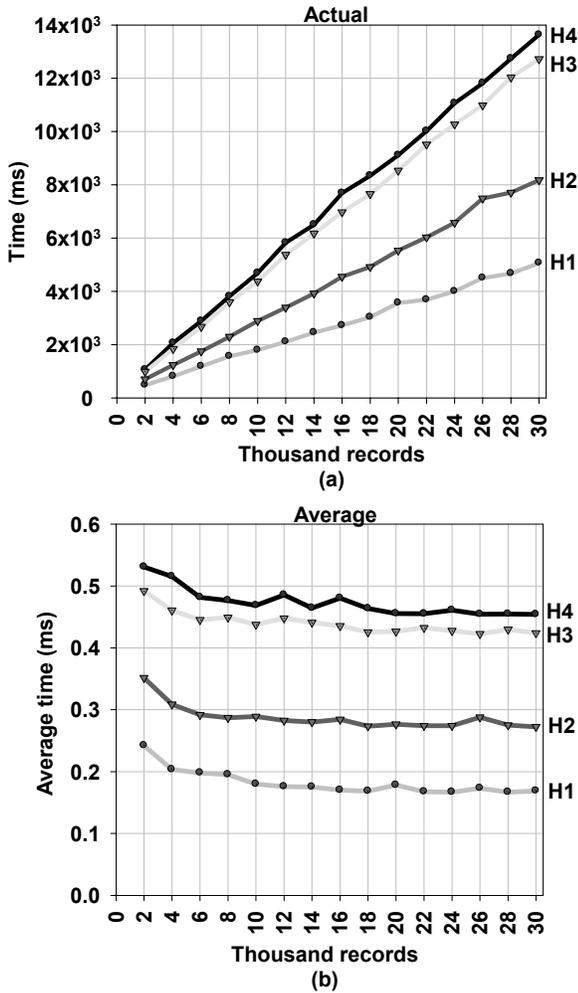

Fig. 5. Algorithm's scalability for 12.5% of data imprecision. (a) Actual time taken by our algorithm with partition trees H1, H3, H4 and H7. (b) Average time to traverse a record by the algorithm for the same trees.

With each type of hierarchy, the average time of defuzzification stabilizes to a constant regardless of the number of records. Based on the charts presented in Fig. 4(b), we are happy to report that the algorithm maintains the linear performance for all investigated partition trees. The defuzzification algorithm shows linear performance even with lower percentage of data imprecision. Exactly the same behaviors are found with 12.5% of data imprecision shown in Fig. 5. Every aspect of Fig. 5 is the same as Fig. 4. Therefore, the experiments conducted in this section show that the average case behavior of the defuzzification algorithm is linear with all partition trees and imprecision levels. We ensured average case by selecting random atomic values from the domain while constructing artificial datasets for different percentage of imprecision.

IV. CONCLUSION

In this work we focus on the problem of interpretation of multi-valued, nonnumeric entries in the similarity-based fuzzy relational databases. The heuristic algorithm presented in this paper allows for transfer of fuzzy records that reflect uncertainty to the form that allows analysis of such data via majority of regular data mining algorithms. Our scalability tests for the developed algorithm show that the fuzzy collections can be transferred to the atomic values in the efficient way. The tests have been conducted with different partition trees and different percentage of imprecision of fuzzy data. Our future goal is to research on the discovery of the most appropriate data defuzzification algorithms.